\documentclass[twocolumn,superscriptaddress,prl, showpacs,floatfix]{revtex4}
\usepackage{epsfig,color}
\usepackage{graphicx}% Include figure files
\usepackage{dcolumn}% Align table columns on decimal point
\usepackage{bm}% bold math
\usepackage{setspace}

\begin{document}

\title{Anisotropic Magneto-conductance of InAs Nanowire: Angle Dependent Suppression of 1D Weak Localization}

\author{Dong Liang, Juan Du, Xuan P.A. Gao}
\email{xuan.gao@case.edu}
 \affiliation{Department of Physics, Case Western Reserve University, Cleveland, OH 44106}

\begin{abstract}
The magneto-conductance of an InAs nanowire is investigated with
respect to the relative orientation between external magnetic
field and the nanowire axis. It is found that both the
perpendicular and the parallel magnetic fields induce a positive
magneto-conductance. Yet the parallel magnetic field induced
longitudinal magneto-conductance has a smaller magnitude. This
anisotropic magneto-transport phenomenon is studied as a function
of temperature, magnetic field strength and at an arbitrary angle
between the magnetic field and the nanowire. We show that the
observed effect is in quantitative agreement with the suppression
of one-dimensional (1D) weak localization.

\end{abstract}
\date{\today}
 \maketitle

In the past two decades, semiconductor nanowires have attracted
much attention due to their great potential in both fundamental
research and novel device applications. They are considered as
versatile building blocks for future nanoelectronics
\cite{Lieber,thelander}, and as a platform to explore the
fundamental physical properties of quasi-1D \cite{Luwei,smallSiNW}
or zero-dimensional (0D) systems \cite{bjork,xiangjie,yongjie,
InAsQDspinorbit, NWholespin, nanowireQD1, nanowireQD2}. Moreover,
the quasi-1D nature of semiconductor nanowires may help to achieve
significant enhancement in thermoelectric performance
\cite{NWTEtheory, NWTE1,NWTE2}. Thus it is highly desirable to
investigate the quasi-1D transport properties of nanowire devices.
Previously, transport studies on nanowire structures have shown
interesting effects related to the quasi-ballistic transport
\cite{Luwei}, quantum confinement\cite{smallSiNW} and interference
\cite{SiNWinterference} effects. Magneto-transport studies also
revealed various quantum transport phenomena such as 1D weak
localization or anti-localization effect
\cite{InAsNWarrayMR,GaNNWarrayMR,simmonsSiNW,dhara,liangdong}.
These results exemplify the wide range of possibilities to study
nanoscale physics in semiconductor nanowire materials made by
direct chemical synthesis.

Here, a comprehensive magneto-transport study is reported on a
single InAs nanowire field-effect-transistor (FET) device. An
anisotropic magneto-conductance is observed with regard to the
orientation between the magnetic field and the nanowire axis. We
show that, this effect can be attributed quantitatively to the
suppression of 1D weak-localization correction to nanowire
conductance. While much work to date has been focused on studying
the magneto-conductance of nanowires in perpendicular
\cite{InAsNWarrayMR,GaNNWarrayMR,simmonsSiNW,dhara,liangdong} or
occasionally, parallel magnetic field\cite{BiNW1,lujia}, we report
the first measurement of the continuous evolution between the
transverse and longitudinal magneto-conductance, and show that the
anisotropic magneto-conductance is explained by the 1D weak
localization model \cite{Altshuler81,beenakkertheory}.

InAs nanowires with 20nm diameters were grown on silicon wafer in
a thermal chemical vapor deposition(CVD) system. Nanowires were
sonicated by ultra-sound sonication and suspended in ethanol which
was then dropped on silicon substrate with a thermal oxide layer
on the surface for device fabrication. The 600nm thick oxide on
silicon substrate was used as the gate dielectric and the highly
doped n-type Si substrate itself was used as the gate electrode.
Ti/Al(2nm/60nm) electrodes with 2$\mu$m spacing were evaporated on
both sides of nanowire to serve as source and drain contacts. The
device was dipped in buffered hydrofluoric acid solution for about
3s before metal evaporation to remove any native oxide and ensure
ohmic contacts. The two-terminal conductance ($G$) of the nanowire
was measured by low frequency lock-in amplifier at a constant
excitation voltage of 1mV. The magneto-transport measurement on
the single InAs nanowire was performed in a Quantum Design
physical property measurement system equipped with a rotating
sample stage. Prior to cool down, the device was carefully aligned
to make sure the axis of nanowire is parallel to the fixed
magnetic field direction (the misalignment is within a few
degrees). Then the computer-controlled rotator was used to rotate
the nanowire to form an arbitrary angle between nanowire and the
the magnetic field $B$ (see Fig.3a).

The dependence of the InAs nanowire conductance $G$ on the gate
voltage $V_g$ at various temperatures and zero magnetic field is
shown in Fig.1. The inset shows the device configuration. The
positive slope of $G(V_g)$ in Fig.1 indicates that the carrier is
$n$-type, similar to reports in the literature
\cite{samuelsonInAs,WangDLInAs,LieberInAs,JaveyInAs,sakrAPL}. In
Fig.1, it can be seen that the $G(V_g)$ curves shift downwards as
$T$ decreases. This reduction of conductance was attributed to the
weak localization correction to the Drude conductance instead of a
decreasing carrier density effect\cite{liangdong}. Below 10K, some
small amplitude oscillations start to develop on top of the
quasi-linear $G(V_g)$ dependence. These oscillations are either
due to the quantization of multiple 1D-subbands with different
mobility \cite{smallSiNW} or interference effects of electron
waves undergoing multiple reflections inside the nanowire channel
\cite{SiNWinterference}. From the trans-conductance $g_m\equiv$
d$G$/d$V_g$, we can estimate the electron mobility \cite{sakrAPL,
dujuan} $\mu=g_m\times L^2/C_g$, where $C_g$ is the gate
capacitance and $L$=2$\mu$m is the length of nanowire between the
electrodes. We can approximate the $C_g$ with a wire on an
infinite plate model and estimate electron mobility
$\mu\approx$205cm$^2$/Vs. The electron concentration can be
roughly estimated as $n(V_g)=(V_g-V_{th})C_g/eL\approx
(V_g-V_{th})\times 280/\mu$m. The threshold voltage $V_{th}$ is
$\sim$ -45V by linearly extrapolating the $G(V_g)$ curve to zero
at $T$=40K when the quantum correction to $G$ is small. Thus for
the range of gate voltage studied (-10 to +10V), electron
concentration ranges from 1.0 to 1.5$\times$10$^4$/$\mu$m, or
equivalently, 3.1 to 4.8 $\times$10$^{19}$/cm$^3$. For such high
electron concentrations, there are many 1D subbands filled,
therefore we use the three-dimensional (3D) formula to estimate
the electron mean free path $l \sim$ 13.1, 14.4 and 15.2nm,  for
$V_g$=-10, 0 and +10V. The corresponding 3D Fermi wavelength
$\lambda_F$ is 6.5, 5.9 and 5.6nm.

\begin{figure}[btph]
\centerline{\psfig{file=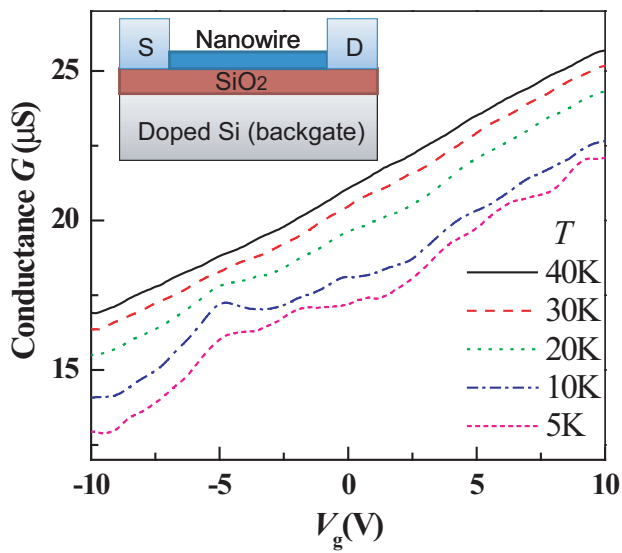,width=8.5cm}}
 \caption{(color online) Dependence of conductance $G$ on gate voltage of
an InAs nanowire at temperature $T$ = 40, 30, 20, 10, 5K. Nanowire
diameter is 20nm. The inset is a schematic of the device which has
2$\mu$m channel length.}
  \label{fig1}
\end{figure}

The magneto-conductance of nanowire is presented in Fig.2a for
both the perpendicular field and parallel field configurations at
several temperatures from $T$=40K down to 5K. There are two main
qualitative observations regarding the magneto-conductance $G(B)$.
$First$, at $B<2$T, both the perpendicular field ($B_\perp$) and
parallel field ($B_\parallel$) induce a positive
magneto-conductance, whose magnitude increases at lower
temperature. The positive magneto-conductance in $B_\perp$ has
been explained by the suppression of 1D weak localization by us in
an earlier paper \cite{liangdong}. Above 2T, the conductance of
nanowire tends to saturate, due to the complete suppression of
weak localization effect\cite{liangdong}. $Secondly$, at all
temperatures, the increment in $G(B_\perp)$ is always larger than
$G(B_\parallel)$. This difference between the transverse
magneto-conductance $G(B_\perp)$ and the longitudinal
magneto-conductance $G(B_\parallel)$ is the main focus of this
paper. Below we will show that both effects arise from the same
mechanism, namely, magnetic field suppression of 1D weak
localization.

Quantum correction to the Drude conductivity in weakly disordered
system at low temperatures stems from weak localization, which is
constructive quantum interference of time-reversal paths formed by
diffusive electrons scattered elastically by impurities or
defects\cite{Bergmann}. Such weak localization effects have been
widely observed in various disordered electronic systems and
received tremendous research interests\cite{LeeRMP}. One key
parameter in this process is the phase coherence length of
electrons $L_\varphi$, indicating the characteristic length within
which the electron maintains its phase coherence for the
interference to take place. Phase coherence can be destroyed by
electron-electron and electron-phonon scatterings. Such inelastic
scatterings are strengthened by increasing temperature, which
leads to the diminishing effect of the weak localization effect at
high temperatures. Moreover, magnetic field can be applied to
introduce additional phase shifts to the electron waves travelling
in time-reversed paths, which suppresses the weak localization and
enhances the sample conductance. The dependence of conductance on
magnetic field is usually used to extract the $L_\varphi$. If
$L_\varphi\gg w$, the width of system, the localization is
regarded as 1D, and the quantum correction to conductance in
magnetic field is given by the following equation
\cite{Altshuler81,beenakkertheory}:
\begin{equation}
\label{eq1}
G(B)=G_0-\frac{2e^2}{hL}{\left(\frac{1}{L_\varphi^2}+\frac{1}{L_B^2}\right)}^{-1/2},
\end{equation}
where $L$=2$\mu$m is the length of the nanowire, $w$=20nm is the
width or diameter of nanowire, and $G_0$ is the classical Drude
conductance. The magnetic relaxation length $L_B=(D\tau_B)^{1/2}$,
with $D$ as the diffusion constant and $\tau_B$ as the magnetic
relaxation time which depends on the field orientation, strength
and wire cross-section shape etc. We first fitted the low field
($B<$1T) $G(B_\perp)$ data in Fig.2a to Eq.1 using
$L_B=L_{B\perp}=\sqrt3\hbar/(eB_\perp\sqrt{\pi (w/2)^2})$
\cite{Altshuler81,beenakkertheory,note}. $G_0$ and $L_\varphi$ are
the only two fitting parameters. The fitting curves are indicated
by the solid line in Fig.2a, while the fitted phase coherence
lengths $L_\varphi$ and the Drude conductance $G_0$ at various
temperatures are shown in Fig.2b. The above expression for
$L_{B\perp}$ is applicable in the so-called 'dirty metal' regime
($l\ll w$), and Eq.1 requires $\sqrt{\hbar/eB}\gg w$ (the weak
field limit). These requirements are satisfied reasonably for our
parameters. As shown in the Fig 2b, $L_\varphi$, ranging from
41.3nm to 103.6nm, is much larger than the diameter $w$ of the
nanowire. This justifies the use of 1D weak localization
($L_\varphi>w$) to fit our data. It is interesting to note that
the extracted $G_0$ has some temperature dependence. We believe
that this residual $T$-dependence in $G_0$ is due to the
electron-electron interaction effects, which also give a
correction to the Drude conductance\cite{simmonsSiNW,lujia}. It is
well known that, unlike weak localization, electron-electron
interaction induced quantum correction to conductance is
independent of magnetic field \cite{LeeRMP}. Thus it will not
affect our study of orientation-dependent magneto-conductance.
\begin{figure}[btph]
\centerline{\psfig{file=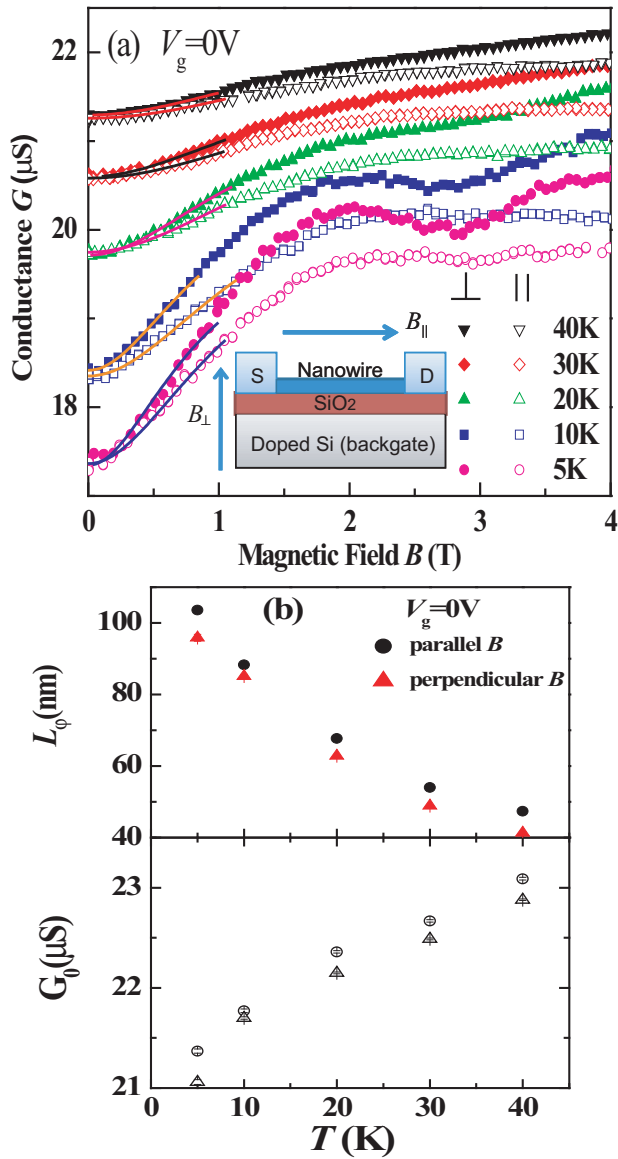,width=8.5cm}}
\caption{(color online) (a)InAs nanowire conductance vs magnetic
field at temperature $T$ = 40,30,20,10,5K in $B_\perp$ (filled
symbols) or $B_\parallel$ (open symbols). The gate voltage
$V_g$=0V. Solid curves show the best fits to 1D weak localization
theory. (b) The electron phase coherence length $L_\varphi$ and
Drude conductance $G_0$ plotted as a function of temperature.
$L_\varphi$ and $G_0$ are obtained by fitting the low field
$G(B_\perp)$ or $G(B_\parallel)$ data to 1D weak localization
theory.}\label{fig2}
\end{figure}

The weak-localization correction to the conductance of a wire in
parallel magnetic field was also developed by Altshuler and Aronov
\cite{Altshuler81}. The analytical expression is the same as Eq.1
but with $L_B\equiv L_{B\parallel}=2\sqrt2\hbar/(eB_\parallel w)$.
We fit the $G(B_\parallel)$ in Fig.2a to Eq.1 using the expression
for $L_{B\parallel}$ and found a good agreement. The extracted
$L_\varphi$ and $G_0$ are shown in Fig.2b to compare with the
results from fitting $G(B_\perp)$ data. There ia a good agreement.
Therefore, both the perpendicular and parallel field induced
magneto-conductance are consistent with the localization effect
with the same parameters. Before we proceed to the angular
dependence of $G(B)$, we like to make a few comments on the
physical origin of the stronger $G(B)$ in the perpendicular
configuration. Evidently, $L_{B\parallel}>L_{B\perp}$, this means
that a perpendicular field is more effective than a parallel field
in suppressing the localization effect. This difference is related
to the 1D size confinement effect of coherent electron waves
($L_\varphi>w$). In the parallel field configuration, electron
diffusion paths enclosing magnetic flux are well confined within
the circumference of the nanowire, while for the perpendicular
field case, they are only confined in the radial direction, and
are free of confinement in the axial direction. So the magnetic
fluxes enclosed by electron diffusion paths in perpendicular field
are larger in the perpendicular field configuration. Thus, the
additional phase shift introduced by a small magnetic field for
$B_\perp$ is larger than $B_\parallel$, which leads to stronger
suppression of weak localization effect.

\begin{figure}[btph]
\centerline{\psfig{file=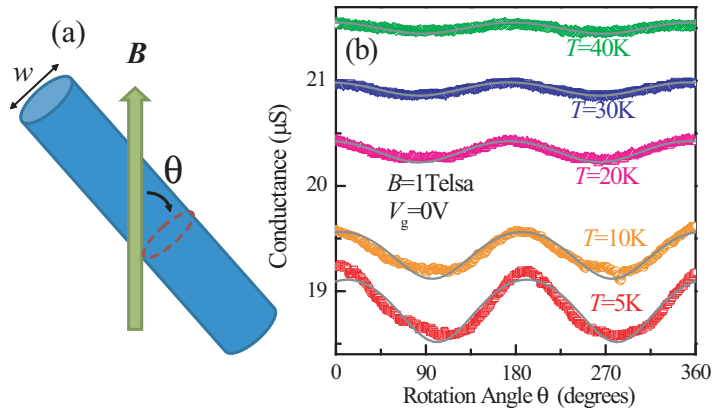,width=9.5cm}}
\caption{(color online) (a) Schematic of rotating a nanowire with
diameter $w$ in magnetic field $B$. Rotation angle $\theta$ is
defined to be the angle between magnetic field and the cross
section of nanowire. (b) Rotation angle dependence of an InAs
nanowire conductance at $V_g$=0V and $T$ = 40, 30, 20, 10, 5K.
Solid curves are the fits to Eq.2. } \label{fig3}
\end{figure}

To further test the conclusion that the anisotropic
magneto-conductance of our InAs nanowire has the same origin of
magnetic field suppression of 1D weak localization, we carried out
an angle dependence experiment of the magneto-conductance at
various temperatures in a fixed magnetic field of 1T. The nanowire
was carefully aligned to make sure it could be rotated about the
axis perpendicular to both the magnetic field and the nanowire, as
shown in Fig.3a. The rotation angle $\theta$ is defined to be the
angle between magnetic field and the cross section plane of
nanowire. The experimental data of angle dependence of
magneto-conductance for $\theta$=0 to 360 degrees are given in Fig
3(b). The gate voltage $V_g$ was set to 0V. According to our
definition of $\theta$, $B$ is exactly perpendicular to the
nanowire at $\theta$=0 and 180 degrees, and is exactly parallel to
nanowire at $\theta$=90 and 270 degrees. Obviously, since that
magneto-conductance in perpendicular field is larger than in
parallel field, $G(\theta)$ data in Fig.3b shows peaks at 0, 180
and 360degrees, and dips at 90, 270 degrees. The small deviations
of the peak and dip positions from 0, 90, 180 and 270 may be
caused by the misalignment of nanowire relative to magnetic field.
In addition, the oscillatory amplitudes of $G(\theta)$ decreases
as the temperature increases because of the weakening of the weak
localization effect at high temperatures. In order to explain the
experimental data at an arbitrary angle $\theta$, we decompose the
magnetic field to a perpendicular component and a parallel
component. These two orthogonal magnetic fields will both
contribute to the suppression of electron localization in
nanowire, but with different magnitudes, as discussed earlier. The
overall magneto-conductance is then given by
\begin{equation}
\label{eq2}
G(B,\theta)=G_0-\frac{2e^2}{hL}{\left(\frac{1}{L_\varphi^2}+\frac{1}{L_{B\perp}^2}+\frac{1}{L_{B\parallel}^2}\right)}^{-1/2}.
\end{equation}
In Eq.2, $L_{B\perp}=\sqrt3\hbar/(eB\cos\theta\sqrt{\pi (w/2)^2})$
and $L_{B\parallel}=2\sqrt{2}\hbar/(ewB\sin\theta)$ are the
magnetic relaxation lengths for the perpendicular and parallel
components of $B$. The solid curves in Fig 3(b) indicate the
theoretical fits to Eq.2, with $G_0, L_\varphi$ fixed at the
averaged values of data in Fig.2b, and a misalignment angle offset
was set as the only free parameter. The fitted angle offset is
within 10 degrees. This good agreement between $G(\theta)$ data
and theory further supports that the anisotropic
magneto-conductance of nanowire is consistently explained by the
1D weak localization theory.

In conclusion, angle dependence of the anisotropic
magneto-conductance of a single InAs nanowire with 20nm diameter
is studied. The data are consistent with the theory of
orientation-dependent suppression of 1D weak localization by a
magnetic field \cite{Altshuler81,beenakkertheory}. Not only will
the results benefit the fundamental understandings of
magneto-transport in nanowires, the continuously tuned
angle-dependent magneto-conductance may open up opportunities to
explore other delicate effects such as different spin-orbit
couplings in semiconductor
nanowires\cite{Nitta2008,Spincontrol09}.

The authors thank L. Qiu and M. MacDonald for useful discussions.
X.P.A.G. acknowledges CWRU startup fund and ACS Petroleum Research
Fund (Grant 48800-DNI10) for supporting this work.


\begin{references}
\bibitem{note} Original equation for the perpendicular field configuration was
obtained for wires with square cross-section in
Ref.\cite{Altshuler81}. We modified the equation to account for
the circular cross-section of nanowire.
\end{references}
\end{document}